\documentclass[11pt]{article}
\usepackage{amsfonts}
\usepackage{graphicx}
\usepackage{epstopdf}
\usepackage{epsfig}
\usepackage{amssymb}
\usepackage{setspace}
\usepackage{caption}
\usepackage{amsfonts}
\usepackage{color}
\usepackage{amsmath}
\usepackage{float}
\usepackage{comment}
\newcommand {\be}{\begin{equation}}
\newcommand {\ee}{\end{equation}}
 \newcommand {\bea}{\begin{array}}
 
 \newcommand {\eea}{\end{array}}

\textwidth=6.5in \hoffset=-.75in \voffset=-1in \numberwithin{equation}{section}
\numberwithin{figure}{section}

\begin{document}

\begin{titlepage}
\vspace{1cm} 
\begin{center}
{\Large \bf {Pair production of scalars around near-extremal Kerr-Sen black holes}}\\
\end{center}
\vspace{2cm}
\begin{center}
\renewcommand{\thefootnote}{\fnsymbol{footnote}}
Haryanto M. Siahaan{\footnote{haryanto.siahaan@unpar.ac.id}}\\
Center for Theoretical Physics,\\ 
Department of Physics, Parahyangan Catholic University,\\
Jalan Ciumbuleuit 94, Bandung 40141, Indonesia
\renewcommand{\thefootnote}{\arabic{footnote}}
\end{center}

\begin{abstract}

We investigate the charged scalar pair production near the horizon of a near-extremal Kerr-Sen black holes. The condition for pair production to occur has relation to the violation of Breitenlohner-Freedman bound in an AdS$_2$ space. The method employed in this work has been used to show the pair production in the near-horizon of a near-extremal Kerr-Newman black hole and its non-rotating case as well. We also discuss the static limit of our result. 

\end{abstract}
\end{titlepage}\onecolumn 
\bigskip 

\section{Introduction}
\label{sec:intro}

Pair production naturally arises when we discuss some quantum aspects of vacuum, either in flat or curved spaces \cite{Wald:1995yp}. In flat space, pair production occurs in an environment with strong electric field referred as the Schwinger effect \cite{Schwinger:1951nm}. An environment with such a strong electric field can also exist near a near-extremal or extremal charged black holes such as Reissner-Nordstrom and Gibbons-Maeda-Grafinkle-Horowitz-Strominger (GMGHS) black holes, including their rotating cases as well. In literature, a series of works \cite{Chen:2012zn,Chen:2014yfa,Chen:2016caa,Chen:2017mnm} have been devoted to show the existence of scalar pair production near the near-extremal black holes in Einstein-Maxwell theory. The authors found a condition resembling the violation of Breitenlohner-Freedman bound in AdS$_2$ for the pair production to occur. This is not too surprising since the geometry of near-horizon black hole under consideration has the AdS factor, namely the AdS$_2\times$S$^2$ for the near-extremal Reissner-Nrodstrom and warped AdS$_3$ for the near-extremal Kerr-Newman. The pair production of scalars is established by presenting the squared Bogolubov coefficients related to such process, obtained by solving the corresponding Klein-Gordon wave equation. In getting the wave solution, one can make use either the inner or outer boundary conditions that lead to the same result.

In an attempt to study black holes in the presence of electromagnetic fields, the Einstein-Maxwell framework is among the simplest approach that one would consider. However, there exist a strong believe that string theory is the answer for quantum gravity search. This yields the black hole solutions coming from its low energy limit \cite{Ortin:2015hya} are worth further studies. The features of these black holes should be contrasted to the Einstein-Maxwell counterparts in order to find some possible discrepancies which might be verified in nature. Nowadays the test can come from real observations, especially after the birth of gravitational wave astronomy \cite{Barack:2018yly} . The particular interest of our present work is the charged and rotating black holes in the low energy of heterotic string theory \cite{Sen:1992ua}, namely the Kerr-Sen black holes. Studies on this black hole keep appearing in the last couple of years \cite{Liu:2018vea,Bernard:2017rcw,Gwak:2016gwj,Huang:2017whw,Uniyal:2017yll,An:2017hby,Lin:2017oag,Peng:2016wzr,Siahaan:2015ljs,Siahaan:2019oik} as a part of strong gravity researches beyond Einstein. In fact, Kerr-Sen black holes are characterized by exactly the same parameters which describe a Kerr-Newman black hole, i.e. mass, charge, and angular momentum. Since both black holes geometry are asymptotically flat, the standard textbook methods in computing these three parameters can apply \cite{Wald:1984rg}. Nevertheless, despite Kerr-Sen and Kerr-Newman black holes are quite similar in many aspects \cite{Ghezelbash:2012qn,Chen:2010zwa}, we learn that they are distinctive in several features \cite{Ghezelbash:2012qn,Bhadra:2003zs}.

Just like for the Kerr-Newman black hole \cite{Chen:2010zwa,Hartman:2008pb}, one can construct the rotating black holes/CFT$_2$ duality in the same fashions as \cite{Guica:2008mu,Castro:2010fd} for Kerr-Sen black holes as well \cite{Ghezelbash:2009gf,Ghezelbash:2012qn}. However, not all the features which exist in Kerr-Newman/CFT$_2$ holography reappear in Kerr-Sen/CFT$_2$ duality. For example the twofold hidden conformal symmetry which presents in Kerr-Newman spacetime \cite{Chen:2010ywa} does not resemble in Kerr-Sen background \cite{Ghezelbash:2012qn}. Nevertheless, the proposal of Kerr-Sen/CFT$_2$ duality in \cite{Ghezelbash:2009gf} gives a strong hint for the existence of pair production near a Kerr-Sen black hole, similar to the Kerr-Newman black hole investigation \cite{Chen:2016caa}. This is because one of the main ingredients in establishing the pair production near a Kerr-Newman black hole is by showing its near-horizon geometry has a warped AdS$_3$ factor, which is the case for Kerr-Sen black hole as well. Recall that this warped AdS$_3$ appearance of the near black hole horizon which allows one to construct the black hole/CFT$_2$ correspondence as proposed in \cite{Guica:2008mu} for Kerr, \cite{Hartman:2008pb} for Kerr-Newman, and \cite{Ghezelbash:2009gf} for Kerr-Sen backgrounds. Motivated by this hint, we investigate the possibility of spontaneous scalar pair production near a Kerr-Sen black hole and its static limit as well.

The organization of this paper is as follows. In the next section, we review some properties of the low energy limit heterotic string theory and its black hole solutions. Section \ref{s.3} is devoted to construct a solution in this theory which associates to the near-horizon of a near-extremal Kerr-Sen black hole. Subsequently, the equation of motion for massive scalar probes in the background is discussed in section \ref{s.4}. Using the radial solution to the equation of motion for probes, Bogolubov coefficients related to the pair production is calculated in section \ref{s.5}. In section \ref{s.6}, we present a holographic point of view for the scattering formula. Finally, we give conclusion in section \ref{s.7}.

\section{Low energy heterotic string theory and Kerr-Sen black hole}\label{s.2}

The effective action in the low limit of heterotic string theory reads \cite{Sen:1992ua}
\be \label{actionSen}
S = \int {d^4 x\sqrt { - g} e^{ - {\tilde\Phi} } \left( {R + \left( {\nabla \tilde\Phi } \right)^2  - \frac{1}{8}F^2  - \frac{1}{{12}}H^2 } \right)} \,,
\ee 
where ${\tilde\Phi}$ is the dilaton field, $R$ is Ricci scalar, $F^2$ is the squared of Maxwell field strength tensor $F_{\mu \nu }  = \partial _\mu  A_\nu   - \partial _\nu  A_\mu $. In the action above, $H^2$ is the squared of 
\be
H_{\mu \nu \lambda }  = \partial _\mu  B_{\nu \lambda }  + \partial _\lambda  B_{\mu \nu }  + \partial _\nu  B_{\lambda \mu }  - \frac{1}{4}\left( {A_\mu  F_{\nu \lambda }  + A_\lambda  F_{\mu \nu }  + A_\nu  F_{\lambda \mu } } \right)\,,
\ee 
where $B_{\mu\nu}$ is an antisymmetric second rank tensor field. Varying the action (\ref{actionSen}) with respect all the incorporating fields yields the equations of motion
\be\label{eom.var.metric}
R_{\alpha \beta }  + g_{\alpha \beta } \left( {\frac{{\left( {\nabla {\tilde\Phi} } \right)^2 }}{2} - \nabla ^2 {\tilde\Phi}  - \frac{R}{2}} \right) = \frac{1}{4}\left( {F_{\alpha \mu } F_\beta ^\mu   + H_{\alpha \mu \nu } H_\beta ^{\mu \nu }  - g_{\alpha \beta } \left( {\frac{{F^2 }}{4} + \frac{{H^2 }}{6}} \right)} \right) \,,
\ee

\be \label{eom.var.dilaton}
\left( {\nabla {\tilde\Phi}} \right)^2  - 2\nabla ^2 {\tilde\Phi}  = R - \frac{{F^2 }}{8} - \frac{{H^2 }}{{12}}\,,
\ee

\be \label{eom.var.vector}
\nabla _\mu  \left( {e^{ - {\tilde \Phi} } F^{\alpha \mu } } \right) = \frac{{e^{ - {\tilde \Phi} } }}{2}F_{\mu \nu } H^{\alpha \mu \nu } \,,
\ee 
and
\be\label{eom.var.Bmn}
\nabla _\alpha  \left( {e^{ - {\tilde \Phi} } H^{\alpha \mu \nu } } \right) = 0\,.
\ee 

Sen \cite{Sen:1992ua} found a set of fields obeying the equations of motion above by performing the Hassan-Sen transformation \cite{Hassan:1991mq} using Kerr metric as the seed solution. Recently, this transformation has been used to map Kerr-Taub-NUT metric in obtaining the Kerr-Sen-Taub-NUT solution \cite{Siahaan:2019kbw} and the rotating C-metric to give the charged, rotating, and accelerating black holes in the low energy limit of heterotic string theory \cite{Siahaan:2018qcw}. In principle, any stationary and axial symmetric solution to the vacuum Einstein equation can be used as the seed metric to obtain a new solution in the low energy limit of heterotic string theory by making use the Hassan-Sen transformation.

In string frame, Sen solution for the line element can be written as
\be \label{metric-stringframe}
ds^2  =  - \frac{{\Sigma \Delta _r }}{{K^2 }}\left( {d{\hat t} - ax^2 d{\hat \phi} } \right)^2  + \Sigma \left( {\frac{{d{\hat r}^2 }}{{\Delta _r }} + \frac{{dx^2 }}{{\Delta _x }}} \right) + \frac{{x^2 \Sigma }}{{K^2 }}\left( {ad{\hat t} - \left( {\Delta _r  + 2m\left(1+\xi^2\right) {\hat r}} \right)d{\hat \phi} } \right)^2 \,,
\ee
where\footnote{In the standard Boyer-Linquist coordinate, $x=\sin\theta$.} $K = {\hat r}^2  + a^2 \Delta _x  + 2m{\xi}^2 {\hat r}$, $\Delta _r  = {\hat r}^2  + a^2  - 2m{\hat r}$, $\Delta _x  = 1 - x^2 $, and $\Sigma  = {\hat r}^2  + a^2 \Delta _x $. This spacetime will be referred as the Kerr-Sen metric and it contains the Kerr-Sen black hole. The non-zero component of $U(1)$ gauge $A_\mu$ and antisymmetric second rank tensor $B_{\mu\nu}$ fields are
\be\label{A-a}
{\rm \bf A} = \frac{{4m{\xi\sqrt{1+\xi^2}}{\hat r}}}{K}\left( {d{\hat t} - ax^2 d{\hat \phi} } \right)\,,
\ee 
and
\be \label{B-ab}
B_{{\hat \phi} {\hat t}}  = \frac{{2am{\xi}^2 {\hat r}x^2 }}{K}\,,
\ee 
respectively. The corresponding dilaton field is
\be \label{dilaton}
{\tilde\Phi}  =  - \ln \left( {\frac{{\Sigma  + 2m{\xi}^2 {\hat r}}}{\Sigma }} \right)\,.
\ee 
Setting $\xi$ to be zero yields all the non-gravitational fields above to vanish, and the theory (\ref{actionSen}) reduces to the vacuum Einstein framework.

It is known that Kerr spacetime possesses the timelike $\zeta^\mu_{\left(t\right)}$ and axial $\zeta^\mu_{\left(\phi\right)}$ Killing vectors which associate to the conserved mass $M$ and angular momentum $J$ of the Kerr black hole. As one would expect, these Killing vectors are inherited to the Kerr-Sen spacetime which then allow one to compute the mass and angular momentum of the black hole. Since the Kerr-Sen spacetime is asymptotically flat, we can employ the standard Komar integral to compute the mass and angular momentum of a Kerr-Sen black hole, i.e. $M = m\left(1+\xi^2\right)$ and $J = am\left(1+\xi^2\right)$ respectively. Note that the definition of rotational parameter for rotating black hole still holds, namely $J=Ma$. Moreover, the charge of Kerr-Sen black holes takes the value $Q =\sqrt{2} m \xi\sqrt{1+\xi^2}$. 

Now let us rewrite the solutions (\ref{metric-stringframe}) - (\ref{dilaton}) in terms of $M$ and $Q$. The spacetime metric expressed in the Einstein frame\footnote{It is the Kerr-Sen line element in Einstein frame which gives the area of the black hole to be an area of a two sphere.} reads \cite{Sen:1992ua,Ghezelbash:2009gf,Peng:2016wzr}
\[
ds^2  =  - \left( {1 - \frac{{2M{\hat r}}}{\Xi }} \right)d{\hat t}^2  + \Xi \left( {\frac{{d{\hat r}^2 }}{{D_r }} + \frac{{dx^2 }}{{\Delta _x }}} \right) - \frac{{4M{\hat r}ax^2 }}{\Xi }d{\hat t}d{\hat \phi}\]
\be\label{metric.Einstein.frame}
~~ + \left( {{\hat r}\left( {{\hat r} + 2b } \right) + a^2 + \frac{{2M{\hat r}a^2 x^2 }}{\Xi }} \right)x^2 d{\hat \phi} ^2 \,,
\ee 
where $\Xi  = {\hat r}\left( {{\hat r} + 2b } \right) + a^2 \left( {1 - x^2 } \right)$, $D_r = {\hat r}\left({\hat r}+2b\right) - 2M{\hat r} +a^2$, and $2b = Q^2/M$. The $U(1)$ gauge, second rank antisymmetric tensor, and dilaton fields have the expressions
\be \label{non.grav.fields}
{\tilde\Phi}  =  - \ln \left( {\frac{\Xi }{\Sigma }} \right)~,~
{\bf A} = \frac{{2 \sqrt{2} {\hat r}Q}}{\Xi }\left( {d{\hat t} - ax^2 d{\hat \phi} } \right)~,~
B_{\hat \phi \hat t}  = \frac{{2 ab \hat rx^2 }}{\Xi }\,,
\ee
where the other components of $A_\mu$ and $B_{\mu\nu}$ vanish. Note that the metric (\ref{metric.Einstein.frame}) describes the non-extremal Kerr-Sen black hole. On the other hand, to show the pair production near the horizon of near-extremal Kerr-Sen black hole, we need first to obtain the near-horizon of a near-extremal consideration of (\ref{metric.Einstein.frame}) together with all the non-gravitational fields companion, especially the U$\left(1\right)$ gauge field.

Our study follows closely the work presented in \cite{Chen:2016caa} where the authors showed that spontaneous pair production of charged scalars can occur near the near-extremal Kerr-Newman black hole. Clearly both Kerr-Newman and Kerr-Sen black holes are quite similar in many aspects, but still they come from two distinct theories so there is no guarantee that all aspects in one case will be resembled in the other. Some distinguishable properties have been reported, for example the stronger gravitational lensing \cite{Bhadra:2003zs}, the non-existence of $Q$-picture hidden conformal symmetry \cite{Ghezelbash:2012qn}, and some gaps in black holes merger estimations \cite{Siahaan:2019oik}. Thus investigating the pair production near a near-extremal Kerr-Sen black hole by following the steps performed for Kerr-Newman case can give rise to a new distinguishable property between the two black holes or just adding another new similarity. For this reason, an investigation on the possibility of pair production near a near-extremal Kerr-Sen black hole is worth to be pursued.

\section{Near-horizon of a near-extremal Kerr-Sen black hole solution}\label{s.3}

The Kerr-Sen spacetime (\ref{metric.Einstein.frame}) may contain a black hole whose outer and inner horizons are located at  
\be \label{horizon}
\hat r_\pm = M-b\pm\sqrt{\left(M-b\right)^2 - a^2}\,.
\ee 
There exist an upper bound of the black hole rotation in order to avoid the existence of a naked singularity. For a Kerr-Sen black hole, the condition is given by $a \le M-b$. Accordingly, its extremality is achieved at $M = a+b$ where the two horizons in (\ref{horizon}) coincide. Some physics related to an extremal Kerr-Sen black holes is conjectured to be holographic dual to a two dimensional CFT \cite{Ghezelbash:2009gf}, where the non-extremal aspects were discussed in \cite{Ghezelbash:2012qn}. Later on, the authors of \cite{Button:2013rfa} extended this duality to the case of near-extremal Kerr-Sen which could have a significant overlapping with the discussion in this paper if only the near-horizon transformation they used is the same with ours. Nevertheless, the dynamics of scalar field perturbations in the vicinity of near-extremal Kerr-Sen black holes has not been discussed in the literature\footnote{A study to show the hidden conformal symmetry of scalar perturbation in the background of extremal Kerr-Sen geometry has been reported in \cite{Siahaan:2014ihe}.}. 

In this section, we obtain the near-horizon geometry of a near-extremal Kerr-Sen black hole which happens to be a part of investigations in \cite{Button:2013rfa}. However, we employ a set of different near-horizon transformation which leads to a distinct outcome. The coordinate transformation reads
\be\label{near.hor.transf}
\hat r \to a + \epsilon r~~,~~\hat t \to \frac{{2a\left( {a + b} \right)t}}{\epsilon }~~,~~\hat \phi  \to \phi  + \frac{{at}}{\epsilon }\,,
\ee 
and the near-extremal condition is given by
\be
M \to \left( {a + b} \right) + \frac{{\epsilon ^2 B^2 }}{{2a}}\,.
\ee 
Applying this transformations to the metric (\ref{metric.Einstein.frame}) and taking $\epsilon \to 0$ yield
\be \label{nearhor.nearextKerrSen}
ds^2  = \Gamma \left( x \right)\left( { - \left( {r^2  - B^2 } \right)dt^2  + \frac{{dr^2 }}{{r^2  - B^2 }} + \frac{{dx^2 }}{{1 - x^2 }}} \right) + \gamma \left( x \right)\left( {d\phi  + rdt} \right)^2 \,,
\ee 
where
\be
\Gamma \left( x \right) = 2a\left( {a + b} \right) - a^2 x^2 \,,
\ee 
and
\be
\gamma \left( x \right) = \frac{{4a^2 x^2 \left( {a + b} \right)^2 }}{{\Gamma \left( x \right)}}\,.
\ee
The accompanying $U\left(1\right)$ vector, dilaton, and second rank antisymmetric tensor fields are
\be\label{A-near}
A =  - \frac{{2\sqrt{2} Qa^2 x^2 }}{{\Gamma \left( x \right)}}\left( {rdt + d\phi } \right)\,,
\ee
\be \label{Phi-near}
\Phi  =  - \ln \left( {\frac{{\Gamma \left( x \right)}}{{a^2 \left( {2 - x^2 } \right)}}} \right)\,,
\ee
and all $B_{\mu\nu}$ components vanish. The vanishing $B_{\mu\nu}$ yields the contribution to $H_{\alpha\beta\gamma}$ comes from the Chern-Simons term only, i.e.
\be \label{H3form-near}
H_{tr\phi }  =  - \frac{{2Q^2 a^4 x^4 }}{{\Gamma \left( x \right)^2 }}\,.
\ee 
These fields altogether solve the equation of motions for $g_{\mu\nu}$, $A_\mu$, $B_{\mu\nu}$, and $\Phi$ derived from the action (\ref{actionSen}). 

Note that in the neutral limit $b=0$, the set of near-horizon transformation (\ref{near.hor.transf}) is similar to that in Kerr-Newman case \cite{Chen:2016caa} at $Q=0$, up to the choice of rotation in $\phi$. This yields the expression of near-horizon metric (\ref{nearhor.nearextKerrSen}) is the same to that in Kerr-Newman spacetime \cite{Chen:2016caa} when the black hole charge is set to vanish, i.e. the resulting metric are both near-horizon of a near-extremal Kerr black hole. Interestingly, the near-extremal parameter $B$ appears in the line element (\ref{nearhor.nearextKerrSen}) only, and again resembles the situation in the near-horizon of near-extremal Kerr-Newman black holes \cite{Chen:2016caa}. Taking the parameter $B\to 0$ in the fields solution (\ref{nearhor.nearextKerrSen}) - (\ref{H3form-near}) above, we recover the extremal Kerr-Sen black holes whose holography was investigated in \cite{Ghezelbash:2009gf}. In fact, the extremal limit of solutions (\ref{nearhor.nearextKerrSen}) - (\ref{H3form-near}) match the global near-horizon fields of \cite{Ghezelbash:2009gf} up to a radial coordinate shift and gauge freedom of the three form field $H_{\mu\nu\kappa}$.  

In Kerr-Newman discussion \cite{Chen:2016caa}, the obtained near-horizon metric reduces to that of Reissner-Nordstrom \cite{Chen:2012zn} after setting $a\to 0$. Clearly, in the static limit, the formulae describing the pair production of scalars near the Kerr-Newman black hole are in agreement to the results for Reissner-Nordstrom black hole. As the matter of fact, the work presented in \cite{Chen:2016caa} can be considered as an extension of their earlier work \cite{Chen:2012zn} where the rotation is incorporated. However, the same approach does not work for Kerr-Sen case in the construction (\ref{nearhor.nearextKerrSen}) since the static limit if this metric is null. This has a later consequence that we cannot established the pair production of scalars near the static GMGHS black hole right from the beginning by imposing $a\to 0$ in the near-horizon metric (\ref{nearhor.nearextKerrSen}). 

\section{Massive scalar probes in a near-extremal Kerr-Sen black holes background}\label{s.4}

In this section we consider a massive charged scalar probe near the horizon of a near-extremal Kerr-Sen black holes.  The corresponding action describing the probe is
\be
S = \int {d^4 x\sqrt { - g} \left( {D_\mu  \Phi ^* D^\mu  \Phi  + m^2 \Phi ^* \Phi } \right)} \,,
\ee 
where $D_\mu   = \nabla _\mu   - iqA_\mu $. Accordingly, the equation of motion from the action above reads
\be\label{KG.gen}
\left( {\nabla _\mu   - iqA_\mu  } \right)\left( {\nabla ^\mu   - iqA^\mu  } \right)\Phi  - m^2 \Phi  = 0\,.
\ee 
Knowing that the near-extremal Kerr-Sen spacetime is stationary, axial symmetric, and asymptotically flat, we can make use of the ansatz 
\be \label{ansatz.probe}
\Phi  = e^{ - i\omega t + in\phi } R\left( r \right)S\left( x \right)
\ee
for the scalar probe. Employing the ansatz (\ref{ansatz.probe}) into eq. (\ref{KG.gen}) yields a set of separable differential equations which can be written as 
\be\label{rad.eq} 
\frac{d}{{dr}}\left( {\left( {r^2  - B^2 } \right)\frac{{dR\left( r \right)}}{dr}} \right) + \left( {\frac{{\left( {\omega  + nr} \right)^2 }}{{r^2  - B^2 }} + \frac{{an\left( {n - 2\sqrt{2} qQ} \right)}}{{\left( {a + b} \right)}} - 2a\left( {a + b} \right)m^2  - \lambda } \right)R\left( r \right) = 0 \,,
\ee
and
\be\label{ang.eq}
\frac{{\Delta _x^{1/2} }}{x}\frac{d}{{dx}}\left( {x{\Delta _x^{1/2} } \frac{{dS\left( x \right)}}{{dx}}} \right) - \left( {\frac{{n^2 }}{{x^2 }} + \frac{{a^2 x^2 \left( {n - 2\sqrt{2} qQ} \right)^2 }}{{4\left( {a + b} \right)^2 }} - a^2 m^2 x^2  - \lambda } \right)S\left( x \right) = 0\,,
\ee 
where $\lambda$ is the separation constant. 

Comparing the radial eq. (\ref{rad.eq}) to the equation of motion for massive scalar probe in AdS$_2$ with mass $m_{\rm eff}$ suggests us to assign an effective mass $m_{\rm eff}$ for the scalar field (\ref{ansatz.probe}) which reads
\be\label{mass.eff}
m_{\rm eff}^2  = m^2  + \frac{{\lambda  - n^2 }}{{2a\left( {a + b} \right)}} - \frac{{n\left( {n - 2\sqrt{2} qQ} \right)}}{{2\left( {a + b} \right)^2 }}\,.
\ee 
Accordingly, the analogous AdS$_2$ radius takes the value
\be \label{RAdS-KerrSen}
R_{AdS}=\sqrt{2a\left( a+b\right)}\,.
\ee 
Unlike in the discussion of Kerr-Newman black hole, the radius (\ref{RAdS-KerrSen}) vanishes as the rotation parameter $a\to 0$. Clearly this deficiency has the same nature as the null result in the metric (\ref{nearhor.nearextKerrSen}) after imposing the static limit.

Regardless the existence of some problems in the static limit of the metric (\ref{nearhor.nearextKerrSen}) and the wave equation (\ref{rad.eq}), let us continue to investigate the pair production near a near-extremal Kerr-Sen black hole. It is known from the study of massive scalar fields in AdS$_2$, there exist a lower bound for the squared of mass $m_{\rm eff}^2  \ge  - \frac{1}{{4R_{AdS}^2 }}$ so the fields  do not posses any instability. Therefore, the violation of this lower bound which reads
\be \label{BFboundGEN}
m_{\rm eff}^2  <  - \frac{1}{{4R_{AdS}^2 }}\,,
\ee 
which according to (\ref{mass.eff}) can be expressed as
\be \label{BFboundEX}
2m^2 a\left( { a + b} \right) + \lambda  - n^2  - \frac{{an\left( {n - 2\sqrt{2} qQ} \right)}}{{\left( {a + b} \right)}} + \frac{1}{4} < 0
\ee 
leads to an instability of scalar fields in the theory. This violation plays a crucial role in the next section where we get to the results showing pair production near Kerr-Sen black hole does exist.

To obtain the Bogolubov coefficients related to the pair production, we need the radial flux of scalar fields which is defined as 
\be \label{Flux}
{\Psi} = i\int {dxd\phi \sqrt { - g} g^{rr} \left( {\Phi D_r \Phi ^*  - \Phi ^* D_r \Phi } \right)} = 2ia \left( {a + b} \right)\left( {r^2  - B^2 } \right)\left( {R\left( r \right)\partial _r R^* \left( r \right) - R\left( r \right)^* \partial _r R\left( r \right)} \right) {\cal W}\,.
\ee 
In the equation above, $\cal W$ is a function which has the form
\be\label{Flux.ang}
{\cal W} = 2\pi \int {\frac{{xdx}}{{\Delta _x^{1/2} }}S^* \left( x \right)S\left( x \right)}  \,.
\ee 
It appears that $\cal W$ in (\ref{Flux.ang}) is exactly the same to that of Kerr-Newman case \cite{Chen:2016caa}, and it has the same value at the two boundaries $r\to 0$ and $r \to \infty$ due to its purely angular dependence. 

\section{Pair production}\label{s.5}

\subsection{Radial solutions}

To compute the vacuum persistence amplitude $\left| \alpha  \right|^2 $ and mean number of produced pairs $\left| \beta  \right|^2 $ for scalars near the black holes, first we need to solve the corresponding equation of motion for the fields. These quantities $\alpha$ and $\beta$ obeys the Bogolubov relation $\left| \alpha  \right|^2  = 1 + \left| \beta  \right|^2 $ and can be expressed as some ratios of the scalar fluxes (\ref{Flux}) in some particular regions. Therefore, let us start from the radial equation (\ref{rad.eq}) which can be expressed in a simpler form
\be \label{rad.eq.simpler}
\frac{d}{{dr}}\left( {\left( {r^2  - B^2 } \right)\frac{{dR\left( r \right)}}{{dr}}} \right) + \left( {\frac{{\left( {\omega  + nr} \right)^2 }}{{r^2  - B^2 }} + {\nu} ^2 } \right)R\left( r \right) = 0\,,
\ee 
where
\be \label{nu2}
{\nu} ^2  = \frac{{an\left( {n - 2\sqrt{2}qQ} \right)}}{{\left( {a + b} \right)}} - 2am^2 \left( {a + b} \right) - \lambda \,.
\ee
The exact solution to the eq. (\ref{rad.eq.simpler}) can be written as
\[
R_h \left( r \right) = C_1 F\left( {i\left( {n + \zeta } \right) + \frac{1}{2},i\left( {n - \zeta } \right) + \frac{1}{2};i\left( {n + \tilde \omega } \right) + 1;\frac{1}{2} - \frac{r}{{2B}}} \right)\left( {r + B} \right)^{\frac{i}{2}\left( {n - \tilde \omega } \right)} \left( {r - B} \right)^{\frac{i}{2}\left( {n + \tilde \omega } \right)} 
\]
\be\label{Solrad} 
+ C_2 F\left( { - i\left( {\tilde \omega  - \zeta } \right) + \frac{1}{2}, - i\left( {\tilde \omega  + \zeta } \right) + \frac{1}{2}; - i\left( {n + \tilde \omega } \right) + 1;\frac{1}{2} - \frac{r}{{2B}}} \right)\left( {r + B} \right)^{\frac{i}{2}\left( {n - \tilde \omega } \right)} \left( {r - B} \right)^{ - \frac{i}{2}\left( {n + \tilde \omega } \right)} 
\ee
where
\[
\tilde \omega  = \frac{\omega }{B}~,~ \zeta ^2  = n^2  + {\nu} ^2  - \frac{1}{4}\,.
\]
Note that the constants $C_1$ and $C_2$ above are complex valued\footnote{In the solution, the hypergeometric function is defined by $	F \left( {\alpha ,\beta ;\gamma ;z} \right) = 1 + \frac{{\alpha \beta }}{{1!\gamma }}z + \frac{{\alpha \left( {\alpha  + 1} \right)\beta \left( {\beta  + 1} \right)}}{{2!\gamma \left( {\gamma  + 1} \right)}}z^2  + \frac{{\alpha \left( {\alpha  + 1} \right)\left( {\alpha  + 2} \right)\beta \left( {\beta  + 1} \right)\left( {\beta  + 2} \right)}}{{3!\gamma \left( {\gamma  + 1} \right)\left( {\gamma  + 2} \right)}}z^3  + \dots $}, and BF bound violation (\ref{BFboundEX}) guarantees the $\zeta$ parameter to be real.

The radius $r = B$ can be considered somehow to act like a horizon in the near-horizon geometry (\ref{nearhor.nearextKerrSen}). However, the spacetime described by the metric (\ref{nearhor.nearextKerrSen}) is already outside the black hole, and in order to maintain the signature of spacetime unchanged we should have $r > B$. In such consideration, we can think of the near region to be $r \to B$, and the asymptotic one is $r \to \infty$. In this set up, we are allowed to expand the solution (\ref{Solrad}) around $r \to B$ in the near region,
\be
R_h \left( r \right) \approx \left( {2B} \right)^{\frac{i}{2}\left( {n - \tilde \omega } \right)} \left( {C_h^{\left( {out} \right)} \left( {r - B} \right)^{\frac{i}{2}\left( {n + \tilde \omega } \right)}  + C_h^{\left( {in} \right)} \left( {r - B} \right)^{ - \frac{i}{2}\left( {n + \tilde \omega } \right)} } \right)\,,
\ee
where we have used $C_h^{\left( {in} \right)}  = C_2~~{\rm and}~~ C_h^{\left( {out} \right)}  = C_1$. On the other hand, the approximation of (\ref{Solrad}) in the asymptotic region $r\to \infty$ requires the transformation for hypergeometric function \cite{Stegun},
\be 
2 F\left( {a,b;c;\frac{1}{2} - \frac{r}{{2B}}} \right) = \left( {\frac{{r + B}}{{2B}}} \right)^{ - a} F\left( {a,c - b;c;\frac{{r - B}}{{r + B}}} \right) + \left( {\frac{{r + B}}{{2B}}} \right)^{ - b} F\left( {b,c - a;c;\frac{{r - B}}{{r + B}}} \right)\,.
\ee 
Equipped with this transformation, we can express the radial solution (\ref{Solrad}) in the asymptotic region as
\be\label{Rinf}
R_\infty  \left( r \right) \approx C_\infty ^{\left( {in} \right)} r^{ - i\zeta  - {\textstyle{1 \over 2}}}  + C_\infty ^{\left( {out} \right)} r^{i\zeta  - {\textstyle{1 \over 2}}} \,.
\ee 
The constants that we are using in the expressions above are
\be \label{CinfIN}
\frac{C_\infty ^{\left( {in} \right)}}{\Gamma \left( { - 2i\zeta } \right)} =   {\frac{{C_1\left( {2B} \right)^{{\textstyle{1 \over 2}} + i\left( {n + \zeta } \right)} \Gamma \left( {1 + i\left( {n + \tilde \omega } \right)} \right)}}{{2\Gamma \left( {{\textstyle{1 \over 2}} + i\left( {n - \zeta } \right)} \right)\Gamma \left( {{\textstyle{1 \over 2}} + i\left( {\tilde \omega  - \zeta } \right)} \right)}}} +   {\frac{{C_2\left( {2B} \right)^{{\textstyle{1 \over 2}} - i\left( {\tilde \omega  - \zeta } \right)} \Gamma \left( {1 - i\left( {n + \tilde \omega } \right)} \right)}}{{2\Gamma \left( {{\textstyle{1 \over 2}} - i\left( {n + \zeta } \right)} \right)\Gamma \left( {{\textstyle{1 \over 2}} - i\left( {\tilde \omega  + \zeta } \right)} \right)}}}\,,
\ee
and
\be \label{CinfOUT}
\frac{C_\infty ^{\left( {out} \right)}}{\Gamma \left( {2i\zeta } \right)}  =   {\frac{{C_1\left( {2B} \right)^{{\textstyle{1 \over 2}} + i\left( {n - \zeta } \right)} \Gamma \left( {1 + i\left( {n + \tilde \omega } \right)} \right)}}{{2\Gamma \left( {{\textstyle{1 \over 2}} + i\left( {n + \zeta } \right)} \right)\Gamma \left( {{\textstyle{1 \over 2}} + i\left( {\tilde \omega  + \zeta } \right)} \right)}}}  +  {\frac{{C_2\left( {2B} \right)^{{\textstyle{1 \over 2}} - i\left( {\tilde \omega  + \zeta } \right)} \Gamma \left( {1 - i\left( {n + \tilde \omega } \right)} \right)}}{{2\Gamma \left( {{\textstyle{1 \over 2}} - i\left( {n - \zeta } \right)} \right)\Gamma \left( {{\textstyle{1 \over 2}} - i\left( {\tilde \omega  - \zeta } \right)} \right)}}} \,.
\ee

Now we can work out the fluxes of our interests using the general flux formula (\ref{Flux}). In the near region we have
\be\label{Flux.in.h}
{\Psi}_h^{\left( {in} \right)}  =  - 4B{\cal W}\left| {C_2 } \right|^2 \left( {a + b} \right)\left( {n + \tilde \omega } \right)a \,,
\ee
\be\label{Flux.out.h}
{\Psi}_h^{\left( {out} \right)}  = 4B{\cal W}\left| {C_1} \right|^2 \left( {a + b} \right)\left( {n + \tilde \omega } \right)a \,,
\ee
and the ones in asymptotic read
\be\label{Flux.in.inf}
{\Psi}_\infty ^{\left( {in} \right)}  =  - 4{\cal W}\left| {C_\infty ^{\left( {in} \right)} } \right|^2 \left( {a + b} \right)\zeta a \,,
\ee
\be\label{Flux.out.inf}
{\Psi}_\infty ^{\left( {out} \right)}  = 4{\cal W}\left| {C_\infty ^{\left( {out} \right)} } \right|^2 \left( {a + b} \right)\zeta a \,.
\ee 
Note that these fluxes obey the conservation condition
\be \label{Flux.relation}
\left| {{\Psi}_{\rm incident} } \right| = \left| {{\Psi}_{\rm reflected} } \right| + \left| {{\Psi}_{\rm transmitted} } \right|
\ee
which correspond to the Bogoliubov relation
\be 
\left| \alpha  \right| = 1 + \left| \beta  \right|\,.
\ee
By comparing the last two equations, we can get
\be 
\left| \alpha  \right|^2  \equiv \frac{{{\Psi}_{\rm incident} }}{{{\Psi}_{\rm reflected} }}\,,\,\left| \beta  \right|^2  \equiv \frac{{{\Psi}_{\rm transmitted} }}{{{\Psi}_{\rm reflected} }}\,.
\ee 

\begin{figure}
	\begin{center}
		\includegraphics[scale=0.6]{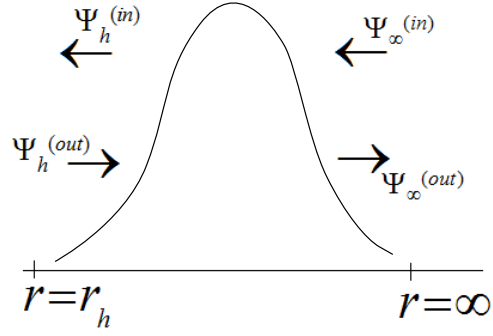}
	\end{center}
	\caption{Illustration of particle's fluxes in the near-horizon of black holes.} \label{Fig.1}
\end{figure}

\subsection{Boundary condition}

As it was proposed in \cite{Chen:2012zn}, one can impose either the inner boundary condition (B.C.), i.e. no outgoing flux at the near region boundary, or the outer B.C. where there is no incoming flux at the asymptotic boundary to get the exact forms of Bogolubov coefficients $\alpha$ and $\beta$. Both conditions lead us to the same results for the vacuum persistence amplitude and mean number of produced pairs. In this paper, we employ the outer boundary condition only, namely by imposing the vanishing incoming flux at spatial infinity, i.e. ${\Psi}_\infty^{\left(in\right)} = 0$. The flux $\Psi _\infty ^{\left( {out} \right)} $ can be interpreted as the particles produced in pair production, repelled electrically by the black hole. On the other hand, the nett of flux $\Psi _h^{\left( {out} \right)} $ and $\Psi _h^{\left( {in} \right)} $ should be interpreted as the beam of antiparticles attracted to the black hole. In such consideration we can borrow an analogy from optical process at an interface, namely 
\be 
\Psi _{\rm incident}  = \Psi _h^{\left( {out} \right)} ~~,~~\Psi _{\rm reflected}  = \Psi _h^{\left( {in} \right)} ~~,~~\Psi _{\rm transmitted}  = \Psi _\infty ^{\left( {out} \right)} \,.
\ee 

The boundary condition $\Psi_{\infty}^{\left(in\right)} = 0$ is fulfilled in (\ref{Flux.in.inf}) if the constants $C_1$ and $C_2$ satisfy
\be 
{C_1} =  - {C_2}{\left( {2B} \right)^{\tfrac{1}{2} - i\left( {\tilde \omega  + n} \right)}}\frac{{\Gamma \left( {1 - i\left( {n + \tilde \omega } \right)} \right)\Gamma \left( {{\textstyle{1 \over 2}} + i\left( {n - \zeta } \right)} \right)\Gamma \left( {{\textstyle{1 \over 2}} + i\left( {\tilde \omega  - \zeta } \right)} \right)}}{{\Gamma \left( {1 + i\left( {n + \tilde \omega } \right)} \right)\Gamma \left( {{\textstyle{1 \over 2}} - i\left( {n + \zeta } \right)} \right)\Gamma \left( {{\textstyle{1 \over 2}} - i\left( {\tilde \omega  + \zeta } \right)} \right)}}\,.
\ee 
Applying the last equation into (\ref{CinfOUT}) gives us 
\be 
C_\infty ^{\left( {out} \right)} =  - {C_2}{\left( {2B} \right)^{\tfrac{1}{2} - i\left( {\tilde \omega  + \zeta } \right)}}\frac{{\Gamma \left( {1 - i\left( {n + \tilde \omega } \right)} \right)\Gamma \left( {2i\zeta } \right)\sinh \left( {2\pi \zeta } \right)\sinh \left( {\pi \left( {n + \tilde \omega } \right)} \right)}}{{\Gamma \left( {{\textstyle{1 \over 2}} - i\left( {n - \zeta } \right)} \right)\Gamma \left( {{\textstyle{1 \over 2}} - i\left( {\tilde \omega  - \zeta } \right)} \right)\cosh \left( {\pi \left( {n - \zeta } \right)} \right)\cosh \left( {\pi \left( {\tilde \omega  - \zeta } \right)} \right)}}\,.
\ee 
As the results of the last two equations, we now have
\be\label{amp.alpha2.gen} 
\left| \alpha  \right|^2  = \frac{{\Psi _{\rm incident} }}{{\Psi _{\rm reflected} }} = \frac{{\Psi _h ^{\left( {out} \right)} }}{{\Psi _h^{\left( {in} \right)} }} = \frac{{\cosh \left( {\pi \left( {n + \zeta } \right)} \right)\cosh \left( {\pi \left( {\tilde \omega  + \zeta } \right)} \right)}}{{\cosh \left( {\pi \left( {n - \zeta } \right)} \right)\cosh \left( {\pi \left( {\tilde \omega  - \zeta } \right)} \right)}}\,,
\ee
and
\be \label{amp.beta2.gen}
\left| \beta  \right|^2  = \frac{{\Psi _{\rm transmitted} }}{{\Psi _{\rm reflected} }} = \frac{{\Psi _\infty ^{\left( {out} \right)} }}{{\Psi _h^{\left( {in} \right)} }} = \frac{{\sinh \left( {2\pi \zeta } \right)\sinh \left( {\pi \left( {n + \tilde \omega } \right)} \right)}}{{\cosh \left( {\pi \left( {n - \zeta } \right)} \right)\cosh \left( {\pi \left( {\tilde \omega  - \zeta } \right)} \right)}}\,.
\ee
Furthermore, the corresponding absorption cross section ratio then reads
\be 
{\sigma _{abs}} = \frac{{\sinh \left( {2\pi \zeta } \right)\sinh \left( {\pi \left( {n + \tilde \omega } \right)} \right)}}{{\cosh \left( {\pi \left( {n + \zeta } \right)} \right)\cosh \left( {\pi \left( {\tilde \omega  + \zeta } \right)} \right)}}\label{sig.abs}\,.
\ee \label{abs.cross.gen}
The non-vanishing of $\left| \alpha  \right|^2$ and $\left| \beta  \right|^2$ leads us to the conclusion that pair production of scalars do exist near the horizon of a near-extremal Kerr-Sen black hole. As one may expect, here we also find ${\left| \beta  \right|^2} =  - {\sigma _{abs}}\left( { - \zeta } \right)$ which is similar to the case of Kerr-Newman \cite{Chen:2016caa}.

Now let us consider the extremal limit $B\to 0$, or equivalently ${\tilde \omega} \to \infty$. As the black hole reaches its extremality, the three quantities above reduces to
\be\label{amp.alpha2.ext}
\left| \alpha  \right|^2  = \frac{{\sinh \left( {2\pi \zeta } \right)}}{{\cosh \left( {\pi \left( {n - \zeta } \right)} \right)}}\exp \left( {n + \zeta } \right)\,,
\ee
\be\label{amp.beta2.ext}
\left| \beta  \right|^2  = \frac{{\cosh \left( {\pi \left( {n + \zeta } \right)} \right)}}{{\cosh \left( {\pi \left( {n - \zeta } \right)} \right)}}\exp \left( {2\pi \zeta } \right)\,,
\ee
and
\be\label{abs.cross.ext}
\sigma _{abs}  = \frac{{\sinh \left( {2\pi \zeta } \right)}}{{\cosh \left( {\pi \left( {n + \zeta } \right)} \right)}}\exp \left( {n - \zeta } \right)\,.
\ee
The non-vanishing of $\left| \beta  \right|^2$ in (\ref{amp.beta2.ext}) distinguishes the Schwinger effect and the Hawking radiation. As in the case of Kerr-Newman black hole, Kerr-Sen black holes cease to Hawking radiate at extremality \cite{Ghezelbash:2012qn}.

To complete the discussions on scalars equation of motion, the angular component (\ref{ang.eq}) can be rewritten as
\be \frac{{{\Delta _x}}}{x}\frac{d}{{dx}}\left( {x{\Delta _x}\frac{{dS\left( x \right)}}{{dx}}} \right) - \left( {\frac{{{n^2}}}{{{x^2}}} + \kappa {x^2} - \lambda } \right)S\left( x \right) = 0\,.\label{ang.eqz}\ee 
The last equation can be transformed to a Heun differential equation type, for example by setting $S\left( x \right) = C_ \pm  x^{ \pm n} H_ \pm  \left( x \right)$. The solution to this angular differential equation is
\be \label{angular.sol}
H_ \pm  \left( x \right){\rm{ = HeunC}}\left( {0, \pm n, - \frac{1}{2},\frac{\kappa }{4},\frac{1}{4}\left( {n^2  + 1 - \lambda } \right),x^2 } \right)\,,
\ee 
where ${\rm HeunC}\left( {\alpha ,\beta ,\gamma ,\delta ,\eta ,z} \right)$ is the confluent Heun function\footnote{See appendix \ref{app.HeunF} for a discussion on this function.}. Nevertheless, since the required calculations related to the pair production are radial dependent only, the angular solution (\ref{angular.sol}) does not need further investigation.

\subsection{Static limit}

The work by Chen et. al. in discussing the possibility of scalar pair production near Kerr-Newman black holes \cite{Chen:2016caa} is an extension of their earlier work \cite{Chen:2012zn} where the black hole under investigation is Reissner-Nordstrom. The Reissner-Nordstrom black hole is electrically charged and non-rotating which can be obtained by setting the static limit $a\to 0$ in the Kerr-Newman solution. Hence, one could expect that imposing the static limit in the vacuum persistence amplitude and mean number of produced pairs near Kerr-Newman black hole yield the corresponding results for Reissner-Nordstrom. It turns out that this is exactly the case. 

In the low energy limit of heterotic string theory, the Kerr-Sen black hole has a static limit known as the Gibbons-Maeda-Garfinkle-Horowitz-Strominger (GMGHS) black hole \cite{Gibbons:1987ps}. However, there exist a difference on the rotating to static transition between the near-extremal Kerr-Newman and near-extremal Kerr-Sen black holes. The static limit of near-extremal Kerr-Newman black hole is the near-extremal Reissner-Nordstrom black hole \cite{Chen:2016caa,Chen:2012zn}. On the other hand, the static limit of the near-extremal Kerr-Sen black hole as in the construction of eq. (\ref{nearhor.nearextKerrSen}) is null. Consequently, one cannot follow the method used in \cite{Chen:2012zn} to show the pair production near a GMGHS black hole in its near-extremality. This situation perhaps can be understood from the fact that a non-extremal GMGHS black hole has a single horizon located $r=2\left(M-b\right)$ instead of the non-vanishing distinctive inner and outer horizons of a generic Reissner-Nordstrom black hole. This could be the reason for the non-existence of $Q$-picture hidden conformal symmetry for Kerr-Sen black hole \cite{Ghezelbash:2012qn}, unlike in Kerr-Newman spacetime \cite{Chen:2010ywa}. 

Despite the null near-horizon metric (\ref{nearhor.nearextKerrSen}) and the vanishing fluxes (\ref{Flux.in.h}) - (\ref{Flux.out.inf}) as the limit $a\to 0$ taken, the squared Bogolubov coefficients (\ref{amp.alpha2.gen}) and (\ref{amp.beta2.gen}) turn out to be non-zero in this limit. Accordingly, we can conclude that the pair production or Schwinger effect is still occurring near the horizon of a near-extremal GMGHS black hole, as it should be for a black hole equipped with large charge. Furthermore, these squared coefficients at extremality (\ref{amp.alpha2.ext}) and (\ref{amp.beta2.ext}) are also not zero as $a\to 0$. Thus one may conclude, according to this outcome, that the pair production still exist for an extremal GMGHS black hole. Recall that the horizons of an extremal GMGHS black hole coincide with its singularity. In the other words, one can think of that the pair production could take place near a GMGHS naked singularity. Nevertheless, the state of extremal GMGHS black hole is impossible to be reached by any physical processes \cite{Horowitz:1992jp}.

\section{Thermal Interpretation}

Following the works presented in \cite{Chen:2012zn,Chen:2016caa,Chen:2017mnm,Kim:2015qma,Kim:2015kna,Kim:2007pm,Chen:2020mqs}, the number of produced particles (\ref{amp.beta2.gen}) can be expressed in terms of instanton actions $S_a  =  - 2\pi n$, ${\tilde S}_a = 2\pi {\tilde \omega}$, and $S_b = 2\pi \zeta$, which reads
\be \label{Nbeta}
{\cal N}  = \left| \beta  \right|^2  = \left( {\frac{{\exp \left( {S_b  - S_a } \right) - \exp \left( { - S_b  - S_a } \right)}}{{1 + \exp \left( { - S_b  - S_a } \right)}}} \right)\left( {\frac{{1 - \exp \left( {S_a  - \tilde S_a } \right)}}{{1 + \exp \left( {S_b  - \tilde S_a } \right)}}} \right)\,.
\ee
Furthermore, in terms of effective temperatures
\be \label{Teff}
T_{{\rm{eff}}}  = \frac{{\bar m}}{{S_a  - S_b }} ~~{\rm and}~~~{\bar T}_{{\rm{eff}}}  = \frac{{\bar m}}{{S_a  + S_b }}\,,
\ee 
the number of produced particles (\ref{Nbeta}) can also be expressed as
\be\label{NinTeff} 
{\cal N} = \exp \left( { - \bar mT_{{\rm{eff}}} ^{ - 1} } \right)\left[ {\frac{{\exp \left( { - \bar mT_{{\rm{eff}}} ^{ - 1} } \right) - \exp \left( { - \bar m\bar T_{{\rm{eff}}} ^{ - 1} } \right)}}{{1 + \exp \left( { - \bar m\bar T_{{\rm{eff}}} ^{ - 1} } \right)}}} \right]\left\{ {\frac{{\exp \left( { - \bar mT_{{\rm{eff}}} ^{ - 1} } \right)\left( {1 - \exp \left( { - \frac{{\hat \omega  - n\Omega _H }}{{T_H }}} \right)} \right)}}{{1 + \exp \left( { - \frac{{\hat \omega  - n\Omega _H }}{{T_H }} - \frac{{\bar m}}{{T_{{\rm{eff}}} }}} \right)}}} \right\}\,,
\ee
where $T_H  = \hat B/2\pi $ and $\Omega_H = \hat{B}$ are the Hawking temperature and angular velocity at the horizon. Here we have used $\hat{\omega} = \epsilon \omega$ and $\hat{B} = \epsilon B$ in denoting the corresponding quantities measured in the original coordinate $\left\{ {\hat t,\hat r,\hat \theta ,\hat \phi } \right\}$ in the spacetime (\ref{metric.Einstein.frame}).

In terms of Davies-Unruh Temperature
\be \label{TU}
T_U  = -\frac{{ n}}{{4\pi \bar ma\left( {a + b} \right)}}
\ee 
and inverse of squared AdS radius (\ref{RAdS-KerrSen}), ${\cal R} = -2/R_{AdS}^2$, the effective temperatures (\ref{Teff}) can take the forms
\be 
T_{{\rm{eff}}}  = T_U  + \sqrt {T_U^2  + \frac{\cal R}{{8\pi ^2 }}}~~~{\rm and}~~~ \bar T_{{\rm{eff}}}  = T_U  - \sqrt {T_U^2  + \frac{\cal R}{{8\pi ^2 }}} \,.
\ee 
The effective mass appearing in (\ref{TU}) is 
\be 
{\bar m} = \sqrt{\frac{1-4 \nu^2}{8 a\left(a+b\right)}}
\ee 
where $\nu^2$ is given in (\ref{nu2}). The terms in (\ref{NinTeff}) can be interpreted as the followings \cite{Chen:2016caa}. The term in square brackets is the Schwinger effect with the effective temperature $T_{\rm eff}$ in AdS$_2$ \cite{Cai:2014qba}, and this effect is associated to the extremal state of the black hole whose near horizon geometry has the AdS$_2$ structure. The contribution of near-extremal state of the black hole to the number of produced particles in eq. (\ref{NinTeff}) is represented by the term in curly brackets, which is considered as the Schwinger effect in Rindler space \cite{Gabriel:1999yz}. It is known that the near horizon geometry of a non-extremal black hole takes the Rindler form. As it is expected, the non vanishing Hawking temperature appears in the Schwinger effect in Rindler space term, since near extremal black hole still Hawking radiates.

Let us make some remarks here. As it has been pointed out in the previous section, Kerr-Sen black holes considered in this paper and Kerr-Newman black holes are quite similar in many aspects, and yet there still exist some differences \cite{Ghezelbash:2012qn,Siahaan:2015ljs,Siahaan:2014ihe}. Here we find another new different features between the two black holes, namely the finiteness of $T_U$ and $T_{\rm eff}$ as one imposes the limit $a\to 0$. In Kerr-Newman case \cite{Chen:2016caa}, taking this limit leads to finite corresponding temperatures, while in Kerr-Sen case yields the singular ones. However, for Kerr-Sen case presented in this paper, the singular effective mass $\bar m$ due to the limit $a\to 0$ yields finite results for the number of produced particles in (\ref{NinTeff}), as it should be as dictated from the original form (\ref{amp.beta2.gen}). The terms ${\bar mT_{{\rm{eff}}} ^{ - 1} }$ and ${ \bar m{\bar T}_{{\rm{eff}}} ^{ - 1} }$ are finite despite $\bar m$, $T_{{\rm{eff}}}$, and ${\bar T}_{{\rm{eff}}}$ diverge individually as one considers $a\to 0$.

\section{Holographic description}\label{s.6}

In addition to performing the semiclassical field in curved space calculations related to the pair production, the authors of \cite{Chen:2016caa} also show that scalar absorption has a two dimensional CFT dual description. Clearly it is correlated to the Kerr-Newman/CFT$_2$ relation proposed in \cite{Chen:2010ywa,Chen:2010zwa}. For Kerr-Sen black hole, such duality has been worked out in \cite{Ghezelbash:2009gf} for the extremal case, and in \cite{Ghezelbash:2012qn} for the non-extremal consideration. Since the hidden conformal symmetry of a non-extremal Kerr-Sen black hole can be established using the $J$-picture only\footnote{Unlike the Kerr-Newman counterpart which has both $J$ and $Q$-pictures \cite{Chen:2010ywa}.} \cite{Ghezelbash:2012qn}, the holography analysis presented in this section is performed in this picture.

Before making a connection between the Schwinger effect for Kerr-Sen black holes and a CFT$_2$ formula, let us recall how a two dimensional CFT can holographically describe the entropy of a non extremal Kerr-Sen black hole. In \cite{Ghezelbash:2009gf}, the author reported that the central charge associated to an extremal Kerr-Sen black reads
\be 
c = 12 J = 12 Ma\,,
\ee 
where $J= Ma$. We then assume that this is also the central charge for the left and right movers CFT$_2$ associated to the geometry of generic Kerr-Sen black hole. Matching the Laplacian in the radial part of test scalar equation of motion with the $SL(2,R)$ squared Casimir yields the left and right temperatures
\be\label{TlTr.ori} {T_L} = \frac{{{r_ + } + {r_ - }}}{{4\pi a}}~~,~~{T_R} = \frac{{{r_ + } - {r_ - }}}{{4\pi a}}\,. \ee
Finally using the Cardy formula coming from the expected two copies of dual CFT$_2$, one can recover the Bekenstein-Hawking entropy for a non-extremal Kerr-Sen black hole
\be {S_{CFT}} = \frac{{{\pi ^2}c}}{3}\left( {{T_L} + {T_R}} \right) = 2\pi M{r_ + } \equiv {S_{BH}}\,.\ee 
From the last equation we can see that a Kerr-Sen black hole could be related to a CFT$_2$ holographically.

Evaluating the left and right temperatures (\ref{TlTr.ori}) into the near-horizon of near-extremal Kerr-Sen background (\ref{nearhor.nearextKerrSen}) gives
\be {T_L} = \frac{1}{{2\pi }}~~,~~{T_R} = \frac{B}{{2\pi a}}\,. \ee
Using these temperatures and the corresponding dual energy excitations in each of the left and right mover theories
\be {{\tilde \omega }_L} = n~~,~~{{\tilde \omega }_R} = \frac{\omega }{a}\,,\ee 
one can find that the general formula for the absorption cross section in a CFT$_2$
\be {\sigma _{CFT}} \sim T_R^{2{h_R} - 1}T_L^{2{h_L} - 1}\sinh \left( {\frac{{{{\tilde \omega }_R}}}{{2{T_R}}} + \frac{{{{\tilde \omega }_L}}}{{2{T_L}}}} \right){\left| {\Gamma \left( {{h_R} + i\frac{{{{\tilde \omega }_R}}}{{2\pi {T_R}}}} \right)} \right|^2}{\left| {\Gamma \left( {{h_L} + i\frac{{{{\tilde \omega }_L}}}{{2\pi {T_L}}}} \right)} \right|^2}\ee 
agrees to the cross section formula computed from gravitational calculation (\ref{abs.cross.gen}), provided that the left and right conformal dimensions
\be\label{hLhR} {h_L} = {h_R} = \frac{1}{2} + i\zeta\,. \ee
Note that unlike the real valued conformal dimensions that appear in the hidden conformal symmetry study \cite{Ghezelbash:2012qn}, $h_L$ and $h_R$ (\ref{hLhR}) are complex valued. This is related to the instability of scalar fields due to the violation of Breitenlohner-Freedman bound to guarantee the existence of pair production near the black hole.

\section{Conclusion}\label{s.7}

Studies presented in this paper are addressed to investigate the pair production in the near-horizon of a near-extremal Kerr-Sen black hole. To do so, we first construct a set of field solutions obeying equations of motion in the low energy limit of heterotic string theory whose metric is related to the near-horizon geometry of a near-extremal Kerr-Sen spacetime. Just like the case of Kerr-Newman black hole, the near -horizon of a near-extremal Kerr-Sen black hole appears to have the warped AdS$_3$ structure. This AdS$_3$ factor plays a crucial role in establishing the pair production near a Kerr-Sen black hole which we present in section \ref{s.5}. Moreover, the absorption cross section related to the scalar pair production can have a CFT$_2$ holographic description. Interestingly, despite the null near-horizon metric and fluxes if the static limit for the Kerr-Sen black hole is considered, applying this limits to the obtained Bogolubov coefficients for Kerr-Sen black hole does not lead to the vanishing results. It tells us that the pair production still exist near a near-extremal GMGHS black hole regardless the lacking of AdS factor in the corresponding near-horizon geometry. 

Our finding that the non-vanishing of vacuum persistence amplitude and mean number of produced pair in the static limit $a\to 0$ is quite surprising, since the near-horizon geometry becomes null in this limit. However, there must exist a way to confirm the pair production near a near-extremal GMGHS black hole, for example in the fashion performed in \cite{Gibbons:1975kk} and \cite{Khriplovich:1999gm}. Or probably there exist an alternative coordinate transformation which suits the spacetime of a near-extremal GMGHS black hole in revealing its AdS$_\times$S$^2$ near-horizon geometry. Provided that this transformation can be found, showing the pair production would be straightforward. We address these works into our future project. 

\section*{Acknowledgement}

This work is supported by LPPM-UNPAR under contract no. PL72018018. I thank my colleagues in the UNPAR Physics Department for their supports. I thank the anonymous reviewer for his/her comments and suggestions.

\appendix
\section{Heun Function}\label{app.HeunF}
This confluent Heun function can be expressed as a polynomial \cite{Marikhin}
\be
{\rm HeunC}\left( {\alpha ,\beta ,\gamma ,\delta ,\eta ,z} \right) = \sum\limits_{i = 0}^k {p_i z^i } \,,
\ee 
where
\be 
c_{i - 1} p_{i - 1}  + a_i p_i  + b_i p_{i + 1}  = 0\,.
\ee 
Variables in the last formula are
\[
a_i  = \frac{{\left( {\alpha  - \beta  - \gamma  + \beta \left( {\alpha  - \gamma } \right)} \right)}}{2} -\eta - i\left( {i - \alpha  + \beta  + \gamma  + 1} \right)\,,
\] 
\[
b_i  = \left( {i + 1} \right)\left( {i + \beta  + 1} \right)\,,
\]
\[
c_i  = \left( {k - i} \right)\alpha \,,
\]
\[
\delta  =  - \left( {k + 1 + \frac{{\left( {\beta  + \gamma } \right)}}{2}} \right)\alpha \,,
\]
and $a_i$, $b_i$, and $c_i$ must satisfy
\[
\det \left[ {\begin{array}{*{20}c}
	{a_0 } & {b_0 } & 0 &  \cdots  & 0  \\
	{c_0 } & {a_1 } & {b_1 } &  \cdots  & 0  \\
	0 & {c_1 } & {a_2 } &  \cdots  &  \vdots   \\
	\vdots  &  \vdots  &  \vdots  &  \ddots  & {b_{k - 1} }  \\
	0 & 0 &  \cdots  & {c_{k - 1} } & {a_k }  \\
	\end{array}} \right] = 0\,.
\]

\end{document}